\documentclass[a4paper,fleqn,usenatbib]{mnras}

\usepackage{newtxtext,newtxmath}
\usepackage[greek,english]{babel}
\usepackage[utf8x]{inputenc}
\usepackage[T1]{fontenc}
\usepackage{ae,aecompl}

\usepackage{graphicx}	
\usepackage{amsmath}	
\usepackage{amssymb}	
\usepackage{xspace}
\usepackage{subfigure}
\usepackage{hyperref}	
\hypersetup{colorlinks=true,linkcolor=blue,citecolor=blue,filecolor=blue,urlcolor=blue}

\newcommand{\gaia}{\emph{Gaia}\xspace}

\title{Exoplanet Transits with Next-Generation Radio Telescopes}

\author[Pope, Withers, Callingham \& Vogt]{Benjamin J. S. Pope$^{1,2}$\thanks{E-mail: benjamin.pope@nyu.edu}, Paul Withers$^{3,4}$, Joseph R. Callingham$^{5}$, and Marissa F. Vogt$^{4}$
\\
$^{1}$Center for Cosmology and Particle Physics, Department of Physics, New York University, 726 Broadway, New York, NY 10003, USA\\
$^{2}$NASA Sagan Fellow\\
$^{3}$Astronomy Department, Boston University, 725 Commonwealth Avenue, Boston, MA 02215, USA\\
$^{4}$Center for Space Physics, Boston University, 725 Commonwealth Avenue, Boston, MA 02215, USA\\
$^{5}$Netherlands Institute for Radio Astronomy (ASTRON), PO Bos 2, 7990 AA Dwingeloo, The Netherlands
}

\date{Accepted XXX. Received YYY; in original form ZZZ}

\pubyear{2018}

\begin{document}
\label{firstpage}
\pagerange{\pageref{firstpage}--\pageref{lastpage}}
\maketitle

\begin{abstract}
Nearly everything we know about extrasolar planets to date comes from optical astronomy. While exoplanetary aurorae are predicted to be bright at low radio frequencies ($< 1~\text{GHz}$), we consider the effect of an exoplanet transit on radio emission from the host star. As radio emission from solar-like stars is concentrated in active regions, a planet occulting a starspot can cause a disproportionately deep transit which should be detectable with major radio arrays currently under development, such as the Square Kilometre Array (SKA). We calculate the radiometric sensitivity of the SKA stages and components, finding that SKA2-Mid can expect to detect transits around the very nearest solar-like stars and many cool dwarfs. The shape of this radiometric light curve will be affected by scintillation and lensing from the planet's magnetosphere and thereby encode magnetospheric parameters. Furthermore, these transits will also probe the distribution of stellar activity across a star's surface, and will help scrub out contamination from stellar activity on exoplanet transmission spectra and radial velocity spectra. This radio window on exoplanets and their host stars is therefore a valuable complement to existing optical tools.
\end{abstract}

\section{Introduction}
\label{intro}

Extrasolar planets have been detected by the transit method with optical telescopes routinely in the two decades since first detection of transits by \citet{henrytransits} and \citet{charbonneautransits}. In this time the transit method has yielded a great amount of information both about planets themselves - including the largest planet survey yields \citep[\emph{Kepler}:][]{2010sci...327..977b}, measurements of planetary radii \citep{2017AJ....154..109F}, and planetary atmospheric characterization \citep{2017haex.bookE.100K} - but also about stars, constraining their mean density \citep{2003apj...585.1038s,2014MNRAS.440.2164K}, limb darkening \citep{2015MNRAS.450.1879E}, and stellar magnetic activity from transits across starspots \citep[e.g. HAT-P-11;][]{hatp11,2018RNAAS...2...26M}. While arguably the first exoplanet to be detected was in the radio, around the millisecond pulsar~PSR\,1257+12 \citep{1992Natur.355..145W}, only one planet around a main sequence star has been detected outside the optical, infrared or UV bands: \citet{2013ApJ...773...62P} detected a deep X-ray transit of an extended atmosphere of the hot Jupiter HD~189733~b in front of its active K~dwarf host star.

Because most main sequence dwarf stars are too radio-quiet to detect with present telescopes, the focus in radio astronomy in relation to exoplanets has been on looking for auroral radio emission from the planets themselves \citep{2000ApJ...545.1058B,2017haex.bookE...9L} or their moons \citep{2014ApJ...791...25N,2016ApJ...821...97N}, which for close-in planets \citep{2015MNRAS.449.4117V}, or for planets around M dwarfs \citep{2018ApJ...854...72T} or giant stars \citep{2007P&SS...55..598Z}, has been predicted to be orders of magnitude more intense than the already bright emission from our own Jupiter \citep{2010ASPC..430..175Z}. The auroral radio emissions are expected to be a sensitive diagnostic of the conditions in the planet's magnetosphere, where by magnetosphere we mean the region of plasma surrounding a planet which is magnetically connected to it rather than to the star or to the background at infinity, and is typically denser than the surrounding interplanetary medium\footnote{In the remainder of this paper, for simplicity, we will refer to a planet's extended plasma environment as its magnetosphere, even if it might properly be referred to as its exosphere or thermosphere. The distinctions in a stratified atmosphere are not important in the simplified models we will be using.}. So far, differential measurements in and out of the secondary transit of the planet passing behind the star have provided only a tentative detection of the aurora of HAT-P-11~b \citep{2013A&A...552A..65L}. Low-frequency radio searches have so far otherwise yielded only upper limits \citep[e.g.][]{2018MNRAS.tmp.1077L}, and recently it has been suggested that self-absorption in dense planetary magnetospheres may make such observations impossible for hot~Jupiters even with planned very large instruments \citep{2017MNRAS.469.3505W,2018MNRAS.479.1194D}. 

There is another possible radio-astronomy approach to characterizing exoplanets, potentially their magnetospheres, and more likely their host star's magnetic configuration: the modulation they cause in the broadband radio flux from the host star. \citet{2018arXiv180906279C} considered the variation imposed by a planet on MHD simulations of coronal radio emission, showing that a planet could cause modulation amplitudes of order unity in the 10-100~MHz frequency regime, and 2-10\% above 250~MHz in some cases. This modulation is caused by the interaction of the planetary magnetosphere with the magnetized stellar corona and wind, which are considered to be smoothly-varying. This is a powerful motivation for low-frequency radio studies of exoplanet host stars, if any are detectable in quiescent corona emission. Here, we wish to consider first the sensitivities of near-future extremely large radio telescopes and evaluate the possibility for detecting this predicted effect, and focus in particular on the effects of planets on smaller spatial scales and correspondingly higher frequencies.

Since the Dover Heights sea-cliff interferometer observations of \citet*{1946natur.157..158p} and \citet*{1947rspsa.190..357m}, it has been known that the most intense solar radio emission comes from sunspots, and subsequent investigations have shown that this is also the case for other main-sequence dwarf stars. These radio emissions can be non-thermal, arising from gyrosynchrotron emission, or thermal. At radio wavelengths the sun's atmosphere attains an optical depth of unity in the corona. Thermal radio emission from these regions therefore has temperatures in thermal equilibrium with the $\sim 10^6$~K plasma \citep{2011SoPh..273..309S}. 

If radio-emitting spots cover only a small fraction of the star's surface, a transiting planet which might only block a tiny fraction of the stellar flux in white light may occult an entire starspot and correspondingly cause a much deeper and briefer radiometric transit. This effect was noted by \citet{2013ApJ...777L..34S} considering ALMA frequencies (hundreds of~GHz), and the contrast between the bright active regions and quiet stellar disk is even more pronounced at cm wavelengths than the submillimetre regime previously considered \citep{2011SoPh..273..309S}. The result is that the broadband geometric transit depth during a spot-crossing event can therefore be very deep ($\sim 10\%$ or more). The depth of the radiometric transit will also constrain spot contamination in transmission spectroscopy, the `transit light source effect' which is currently a major limiting factor in how well atmospheric composition is constrained by optical observations \citep{2015MNRAS.448.2546B,2017LPICo2042.4032R,2018arXiv180202086Z}. Many stars are now known to host planets which frequently transit starspots. These spot-crossing events appear as a brightening anomaly during planetary transits, and have been noted in more than ten systems, such as CoRoT-2 \citep{corot2}, Kepler-17 \citep{kep17}, HD~189733 \citep{2007A&A...476.1347P}, and HAT-P-11 \citep{hatp11,2018RNAAS...2...26M}. Such events have been used to constrain the misalignment of planetary orbits and stellar rotations \citep{2011ApJ...740L..10N}, and measure the spatial distribution of activity across a stellar surface and temporally over the stellar activity cycle.

In this paper, we consider the detectability of stars with forthcoming radio telescopes, finding that neither ALMA nor low-frequency telescopes are likely to detect quiescent radio emission from single main-sequence stars. We wish to extend the proposal of \citet{2013ApJ...777L..34S} to~GHz frequencies, where there are better hopes for detecting transits, and where we will show that interesting studies of exoplanet magnetospheres and of stellar activity are possible. The sharp features of star spots at these frequencies allow for significant modulation on short timescales not previously considered by \citet{2018arXiv180906279C} and with greater sensitivity than is possible at ALMA. 

Stars at these frequencies will be accessible with the Square Kilometre Array \citep[SKA;][]{2009IEEEP..97.1482D}, a large radio interferometer project which will be split between a low-frequency ($\sim$50-350~MHz) arm `SKA-Low' centred on the Murchison Radio Astronomy Observatory near Boolardy, Western Australia, and a higher-frequency ($\sim$350~MHz-14~GHz, noting that not all of the wavebands in that range are fully funded) arm `SKA-Mid' centred on the Karoo Desert in the Northern Cape province of South Africa, with stations throughout southern Africa. The SKA will be built in two stages\footnote{unitedkingdom.skatelescope.org/ska-project/ska-timeline/}, a first-phase SKA1 to be completed around 2023, and a complete second-phase SKA2 to be completed $\sim$~2030.
Although radio observations have been possible only for a handful of unusually active stars with present technology, or for the very closest inactive stars \citep{2014ApJ...788..112V,2018ApJ...857..133B,2018arXiv180308338T}, the SKA will increase our sensitivity by orders of magnitude, and enable the detection of activity-modulated radio emission in more distant and more magnetically-normal stars. We discuss the detailed sensitivities of each instrument and phase in Section~\ref{observing}. At its highest frequencies and longest baselines the SKA will have an angular resolution of $\sim2$~mas, comparable to the typical range of angular diameters of nearby radio emitting solar-like stars \citep[e.g. $\tau$~Ceti and $\epsilon$~Eridani with limb-darkened angular diameters of $2.078  \pm 0.03$~mas and $2.148  \pm 0.03$~mas respectively:][]{2004A&A...426..601D}. While this is sufficient for imaging red giants and for constrained model fits to diameters of main-sequence stars to tens of parsecs \citep{2018arXiv181005055C}, planetary transits will resolve much finer surface features than otherwise possible at radio frequencies.

Transits with the SKA will probe planetary magnetospheres for the first time as they are back-lit by compact, bright stellar active regions. Under favourable conditions, refraction and scintillation through the transiting planet's magnetosphere will modulate the radio light curve in ways that are diagnostic of the magnetic and plasma conditions in the planet's local environment, complementary to the low-frequency auroral emission. 

In the following sections, we will go through simple calculations to illustrate each of these effects and describe their value. There is at present limited literature on radio properties of Sun-like stars, and our calculations are accordingly crude. It will be valuable both to have improved simulations to predict these effects more accurately, and to constrain what can presently be constrained by observations of the Sun and other nearby stars with present-day instruments. 

In Section~\ref{observing} we discuss the stars, instruments, frequencies, and time samplings suitable for radio transit 
studies. In Section~\ref{tess_sec} we assess the likely impact of the Transiting Exoplanet Survey Satellite (TESS) mission on the potential of radio transit observations.
In Section~\ref{geometrictransit} we discuss how the obstruction of stellar radio emissions by a transiting exoplanet affect the observable radio emissions, noting the possibility of using such observations to trace stellar activity and considering how stellar activity affects transit spectroscopy.
In Section~\ref{mdwarfs} we consider the unique characteristics of M dwarfs, which are common hosts for transiting exoplanets and which have relatively unusual radio emissions.
In Section~\ref{mag} we discuss how propagation of stellar radio emissions through the extended magnetosphere of a transiting exoplanet affect the observable radio emissions, considering lensing and scintillation for a Jupiter analogue and a hot Jupiter. We also consider how starspot group size impacts these effects and whether starspot distribution can be inferred from them.

\section{Detecting Stars in the Radio}
\label{observing}

We first need to establish which stars, instruments, frequencies, and time samplings it will be possible to use for radio transit studies. We consider several plausible stellar sources. While radio emission is reported from young T~Tauri stars and interacting binaries, these are not necessarily the most interesting objects of study from the perspective of exoplanetary science and we restrict ourselves to single or well-separated main-sequence stars in this work. 

The Sun was the first bright astronomical radio source, discovered independently by \citet{1944ApJ...100..279R,alexander,army} and \citet{1946natur.157..158p} \citep[as discussed in][]{2006JAHH....9...35O}. Thermal radio emission has been detected in only a handful of solar-like stars: both components of $\alpha$~Centauri \citep[G2V + K1V;][]{2018arXiv180308338T}; $\tau$~Ceti (G8V), $\eta$~Cassiopeiae~A (F9V) and 40~Eridani~A \citep[K0.5V;][]{2014ApJ...788..112V}; and $\epsilon$~Eridani \citep[K2V;][]{2018ApJ...857..133B}. As they resemble the Sun, these are from an astrobiological perspective the most interesting stars as exoplanet hosts, and the bulk of this paper will consider solar-like stars. Of these, we use $\epsilon$~Eri as the standard for GHz-frequency observations, as at its high activity level  it is a better analogue for the very active stars that are of primary interest as spot-crossing transit hosts. $\epsilon$~Eri has an X-ray luminosity of $10^{28.5} \text{erg s}^{-1}$ \citep{1981ApJ...243..234J,1995ApJ...450..392S} compared to the Sun's $10^{26.8}-10^{27.9} \text{erg s}^{-1}$, which is expected to correlate with radio emission \citep{2018ApJ...857..133B}. 

Very significant radio emission is known in M/L~dwarfs \citep{2017arXiv170704264W}: in Section~\ref{mdwarfs} we will discuss the ways in which radio transit observations may give us a better understanding of the unusual magnetospheres of cool and ultracool dwarfs. These are however not typically hosts of the massive hot~Jupiters which in Section~\ref{mag} we will see are some of the most exciting prospects for radio transit observations. 

The radio emissions of single early-type main sequence stars poorly understood, but very few transiting planets are known so far around them either \citep[e.g.][]{2010MNRAS.407..507C,2011AJ....142..195S,2015AJ....150...12B,2015AJ....150..197H,2018A&A...612A..57T,2018AJ....155...35S}. We therefore exclude these from the discussion in the remainder of this paper, while noting that these may be a valuable subject for future work. 

The \citet{2013ApJ...777L..34S} proposal has not so far yielded a submillimetre transit detection due to the intrinsic faintness of the target stars in the ALMA band: the most promising nearby active stars with transiting planets typically have WISE flux densities at 13.6~THz of $\sim$~a few mJy, which is typically the lowest frequency at which any of the aforementioned spot-crossing planet hosts have been detected. Their fluxes fall steeply to longer wavelengths: modelling these sources as black-bodies, the Rayleigh-Jeans law (spectral radiance $\propto \text{frequency}^2$) implies that these would typically be $10^4$ times fainter at the $\sim 100$~GHz frequencies at which ALMA is most sensitive, or around 100~nJy. To detect a 1\% transit with a signal to noise ratio (SNR) of 3 we would therefore need to reach a sensitivity of 0.3~$\mu$Jy on the $\sim$~5~min timescale relevant to spot crossings. Using the ALMA sensitivity calculator, we find that using the full 43~antenna 12\,m~array operating at the highest available frequency (950~GHz) over the widest possible bandwidth (8~GHz, dual-polarisation), the root-mean-square point source sensitivity of a 5~min integration is only 3.4~mJy, a noise floor four orders of magnitude higher than the signal required; even over a full 3~h integration we reach only 0.5~mJy sensitivity, too poor to detect the stars themselves, let alone a transit. Our calculation is more pessimistic than \citet{2013ApJ...777L..34S} in that we demand an SNR of 3, rather than unity; we wish to have a time cadence of 5~min rather than 1~hr, in order to resolve the shape of the light curve; and we consider known spot-crossing transiting planets which are typically more distant than the previously-considered standard of the Sun at 10~pc. 

We therefore extend the analysis to lower radio frequencies, such as will be achievable with the SKA. Using the specifications in the SKA memo SKA-TEL-SKO-0000818 \citep{memo}, we calculate the thermal noise levels in $\mu$Jy for phase-1 SKA1-Low and SKA1-Mid, the second-phase SKA2-Low and SKA2-Mid, and for the existing NSF Karl G. Jansky Very Large Array \citep[VLA;][]{vla}. The planned successor to the VLA, the the Next-Generation Very Large Array \citep[\emph{ng}VLA;][]{ngvla}, which is planned to begin operations from 2028-2034\footnote{\url{http://ngVLA.nrao.edu/page/faq}}, will extend to frequencies above SKA-Mid and below ALMA, as high as 116~GHz, and will have a sensitivity comparable to that of SKA2-Mid in the shared frequency range ($\approx$\,0.2--0.3\,$\mu$Jy between 3 and 20\,GHz for a similar bandwidth and one-hour integration), but its design is still in such a state of flux that the values here should be considered very preliminary. 

\begin{figure*}
\noindent\includegraphics[width=15cm,keepaspectratio]{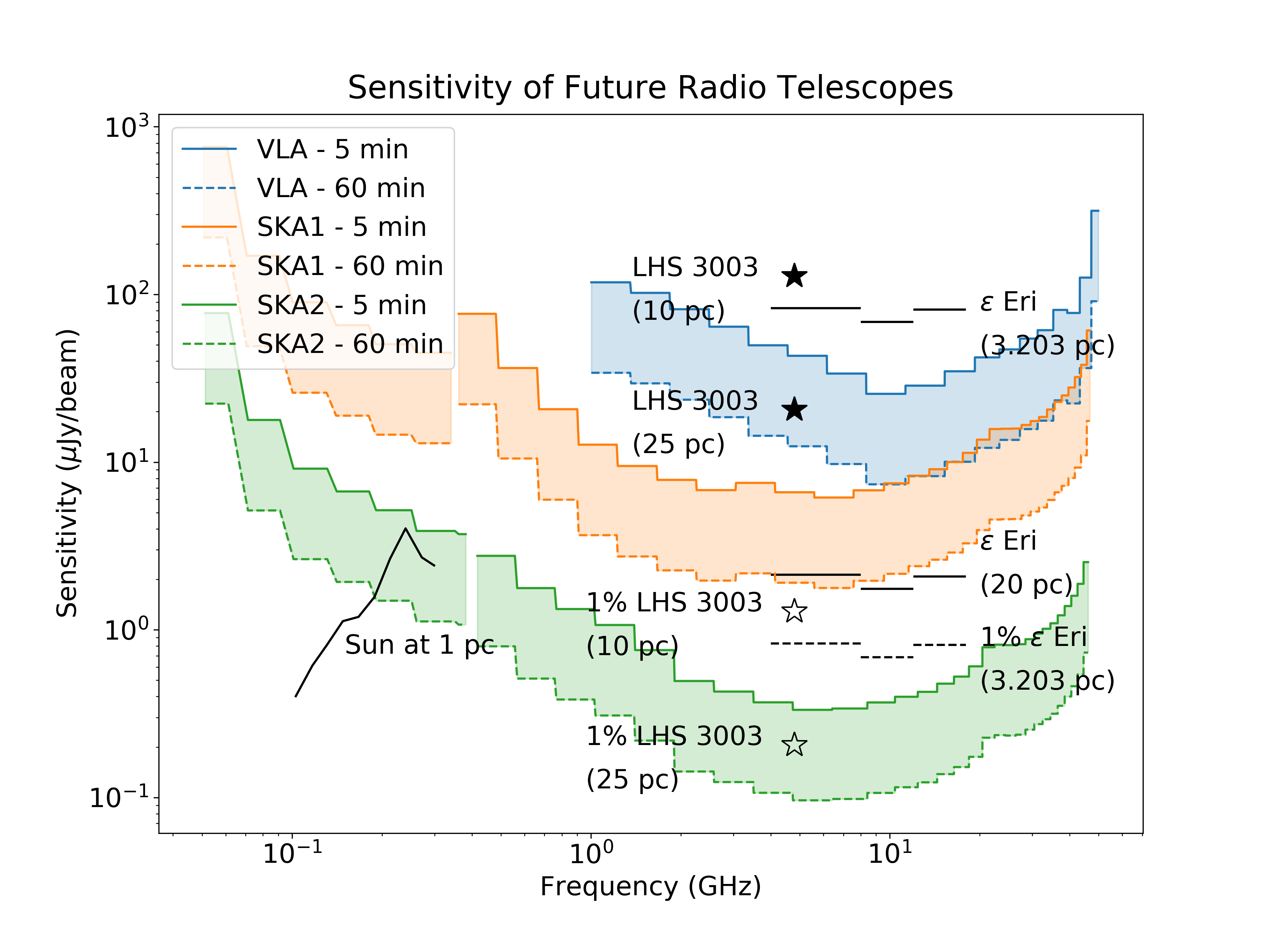}

\caption{\label{sensitivities}
Plot of the $3~\,\sigma$ noise levels of future radio telescopes, assuming no calibration errors and natural weighting, as a function of frequency and for different integration times. At the top with the lowest sensitivity (blue), we have the VLA; below (orange) SKA1; and at the bottom with the highest sensitivity (green), SKA2. The break between the low and high frequency sections is the break between SKA-Low and SKA-Mid, while the high frequency end of VLA is truncated to consider only those portions overlapping with SKA-Mid. The solid lines denote the desired integration time of 5~min, at which features of the transit such as spot crossing events might be detected, and the dotted lines integration times of 1~hr, which might be used for a bare-bones detection. The low-frequency spectral energy distribution (SED) of the Sun at 1~pc \citep[taken from][solid line]{mwasun}, the broadband emission of $\epsilon$~Eri \citep{2018ApJ...857..133B} (solid line segments) at its true distance of 3.203~pc, and of LHS~3003 \citep[taken from][]{2005ApJ...626..486B} at 10~pc and 25~pc (filled star symbols), and $1\%$ of $\epsilon$~Eri (dotted lines) and LHS~3003 (empty star), are shown as indicative targets. At the limits of its sensitivity SKA1-Mid may detect emission from solar-like stars at 20~pc, while SKA2-Mid would detect 1\% transits on  $\epsilon$~Eri itself, or equivalently its emission at ten times its distance (32~pc). Meanwhile SKA-Low in either incarnation is unlikely to detect quiescent emission from any but the very nearest solar-like stars, and will very probably not be useful for transit work. The situation is much more optimistic for late-type stars: transits on LHS~3003 could be detected out to 25~pc, giving us hope for detecting planetary transits around many M~dwarfs.
}
\end{figure*}

In Figure~\ref{sensitivities}, we show thermal noise limited $3~\sigma$ sensitivity levels for both 5~min and 60~min integrations as a function of frequency for each instrument. We note that these values do not include any calibration or instrumental errors, representing an idealized observation, and assume natural weighting. The bandwidth integrated over to reach the sensitivities plotted in Figure~\ref{sensitivities} was chosen noting that the intrinsic spectrum of the observed source is convolved with the bandpass of the instrument. While the radio spectra of the potential targets are relatively unconstrained, we do not expect their spectra to be flat ($\alpha \sim 0$, where the flux density $S_{\nu}$ at frequency $\nu$ is given by $S_{\nu} \propto \nu^{\alpha}$). If any of the sources observed have steeply varying spectral energy distributions, integrating over too large a fractional bandwidth can in fact lead to a decline in sensitivity since more noise than signal is being averaged. To minimise this error in deriving the sensitivities for the various telescopes, we chose to limit the integrated bandwidth to no more than 30\% of the available bandwidth. The total integrated bandwidth was also limited to 2 and 4 GHz for SKA-mid and VLA, respectively. We note that the SKA1-mid bands sensitive between 2-5 GHz and 11-13 GHz are not part of the current planned deployment.

We overplot in Figure~\ref{sensitivities} the flux densities and frequencies of our toy-model targets: $\epsilon$~Eridani with its 4-18~GHz radio fluxes as measured by \citet{2018ApJ...857..133B}, and the same fluxes scaled from the calibrated \emph{Gaia}~DR2 distance from \citet{gaiadistances} of $3.203 \pm 0.005$~pc (\gaia DR2 source ID: 5164707970261630080) to distances of 20~and 40~pc as an analogue to the kind of solar-like targets we might expect to observe. Likewise radio fluxes of the M~dwarf LHS~3003 from \citet{2005ApJ...626..486B} are taken as a fiducial M~dwarf target, which we place at~10 and~25 pc. For each of these we consider a canonical 1\% transit, representing a giant planet transiting a solar-like star, or given their smaller radii, a terrestrial planet transiting an M~dwarf. As no quiescent emission has been detected from solar-like stars at low frequencies, we scale from the solar radio spectrum reported by \citet{mwasun} using the MWA, plotting a solar analogue at~1~pc. 

As is evident in Figure~\ref{sensitivities}, it is barely possible to just detect the unmodulated emission of an $\epsilon$~Eri analogue out to 20~pc with the SKA. This would not be sufficient to yield a transit or to detect coronal emission modulation such as proposed by \citet{2018arXiv180906279C}. The science goals laid out in this paper are achievable only with the full SKA2 (and likely the \emph{ng}VLA): it would be possible with SKA2-Mid using 5~min integrations to detect a 1\% dip out to a few pc, such as around $\epsilon$~Eri should it host any transiting planets. Nevertheless for M~dwarfs, as is evident in the figure, it would be possible to attain 1\% photometry in the requisite timescales out to $\sim 25$~pc with SKA2-Mid, meaning that these are very compelling candidates for such studies. 

On the other hand, with the SKA-Low in either phase, only an extremely nearby solar analogue would be detectable at all, and a transit detection around a solar analogue would certainly be impossible to detect. Prospecting for auroral emission remains the most promising method for investigating the properties of exoplanets at these low frequencies. As no solar-like star has been detected at low frequencies with existing instruments, it is difficult to constrain the radio brightnesses of stars other than the Sun, but in the event that more active stars are significantly more luminous at these frequencies it is possible that active stars may be detected out to a few parsecs.

For planetary transits at the limits of instrumental sensitivity we may want to integrate for longer than is achievable in one orbit. If the period of the orbit is known, we recommend folding and stacking the data according to the known ephemerides. The planetary signal averaged over many orbits then traces the convolution of the mean underlying spot distribution and planet response function. This will be especially feasible for short-period planets and hot Neptunes and Jupiters generally.

Because both components of the SKA are or will be located in the south, northern targets, and in particular the \emph{Kepler} field, will not be observable. The SKA sites, centred at latitudes of $\sim -30^\circ$, strictly cannot observe targets north of $+50^\circ$, and in practice would be restricted to targets some twenty degrees south of this. 
The \emph{ng}VLA in North America will have a complementary field of view capable of observing all targets unobservable with the SKA, provided it reaches the planned SKA2-comparable sensitivity. 
Circumpolar stars (which for the SKA means stars within $\sim 30^\circ$ of the South Celestial Pole) will be continuously observable in the radio; we suspect that the most valuable targets will be those near the Southern Ecliptic Pole in the continuous viewing zone of TESS and the James Webb Space Telescope \citep[JWST:][]{2006SSRv..123..485G}, which will be simultaneously observable with radio and optical instruments for extended durations ($\sim 1$~yr for TESS hemispheres).

\subsection{Finding Targets}
\label{tess_sec}

As we have seen above, it is impractical to detect radio transits, even with SKA2, except across the very nearest stars, around which no transiting planets are currently known. 
The recent launch of the TESS photometric space mission \citep{tess} raises the prospect of discovering nearer candidates with better spot-crossing light curves which will be more appropriate for this study than the stars already known with spot crossing planets. TESS will cover nearly the entire sky in a succession of 27~d, $24\times96$~degree field of view Sectors, beginning with a one-year campaign on the southern celestial hemisphere before continuing for another year in the northern hemisphere, covering ecliptic latitudes greater than $6^\circ$ and overlapping at the ecliptic poles to allow for a continuous viewing zone. It is optimized to detect transits around bright stars, including naked-eye stars. \citet{2018arXiv180405050B} predict that TESS will detect 47~transiting planets within 50~pc, finding the closest such planet in their simulations at 2.7~pc away. At such a close distance, it would be possible to detect a 1\% radio transit with SKA2, whether from a giant planet or from a terrestrial planet occulting a spot. TESS has already delivered its first planet candidates: while at 18.27~pc the naked-eye TESS transit hosting G0 dwarf $\pi$~Mensae \citep{pimensae,pimensae2} is too far away to expect to detect in the radio, TESS has also discovered a hot super-Earth around the nearby (15~pc) M~dwarf LHS~3844 \citep{2018arXiv180907242V}, an ideal candidate for SKA followup.

In addition to TESS, the SPECULOOS ground-based transit survey \citep{speculoos} will observe $\sim 1200$ ultracool dwarfs out to $\sim 50$~pc with spectral type M7. It is projected to discover $\sim 40$ planets in $\sim 20$ such systems.  We hope that TESS and ground-based M~dwarf-targeting surveys such as MEarth \citep{mearth,2015csss...18..767I} and NGTS \citep{ngts} will deliver many more nearby targets for radio transit measurements, especially hot~Jupiters whose magnetospheres might be probed. Similarly, it will be important to crossmatch surveys with the SKA precursors with \gaia to search for radio emission from deep, complete catalogues of nearby late type stars to identify the most promising objects for follow-up (Callingham et al., in prep.). These targets, as discussed in Section~\ref{mdwarfs}, are easier to detect in the radio than hotter stars, and will be valuable targets for radio characterization. 

\section{Broadband Geometric Transit}
\label{geometrictransit}

The main effect likely to be observed is the geometric planetary transit, in which the body of the planet occults part of the star and blocks out its radiation, as was first discussed by \citet{2013ApJ...777L..34S}. In this Section we will consider the effect of a planet transiting the star in broadband radio, ignoring diffraction and refraction. Planets (with diameters of many thousands of kilometres) are much larger than the $\sim$~cm wavelength of commonly used radio instruments, 
so diffraction may be neglected.  

In Section~\ref{mag} we will consider the 
effects of refraction
on radio propagation in the inhomogeneous plasma environment of a planet's magnetosphere.
The broadband radio light curve was investigated by \citet{2013ApJ...777L..34S}, who used Nobeyama Radioheliograph maps of the Sun at 17~GHz to simulate the appearance of an exoplanetary transit at sub-mm wavelengths. They noted two main differences compared to optical transits: an enhanced transit depth crossing stellar active regions, and the change from limb-darkening to limb-brightening which alters the transit shape.

Ignoring stellar activity for the moment, the whole stellar corona is a source of radio waves and the star has an accordingly larger radius \citep[a $\sim 100$~arcsec increase in radius at $\sim$~GHz frequencies, in the case of the Sun;][]{1979ApJ...234.1122K,1996ASPC...93..387G}, which will cause the radiometric transit ingress to slightly precede that of the optical transit, and similarly delay the egress. Timing this effect precisely gives a direct measure of the extent of the stellar chromosphere or corona (probing different regions depending on wavelength); depending on the duration of transit and achievable time sampling, this may be a challenging but worthwhile observation. Likewise, the baseline transit shape will differ significantly from the optical because where in the optical regime the Sun and sun-like stars are limb-darkened, they are actually limb-brightened at radio and millimetre wavelengths \citep{1979ApJ...234.1122K,1981ApJ...244..340H,2016ApJ...825..138D}, so that instead of having a ``u'' shape as in the optical, the transit will actually have a ``w'' shape in the radio. Exoplanetary transits are the only way of detecting this effect in single stars more distant than the Sun without resolving them on extremely long baselines, a key observation for models of stellar chromospheres \citep{2011ApJ...734...64S}. In this regime transit depths for a given star-planet pair are expected to be slightly shallower than in the optical case, owing to the increased stellar diameter.

Beyond transits of the quiescent stellar disk, the Sun is highly inhomogeneous in its distribution of radio flux, with active regions (i.e. starspots) having radio surface brightness orders of magnitude greater than the quiescent disk. 
The extent of the stellar surface covered by such regions is a matter of interest from optical transit studies \citep[e.g.][]{2017ApJ...834..151R,hatp11}, and typical values may range from a few percent to a few tens of percent. As a result, exoplanet transits of active stars are likely to have sharp, brief, deep dips as the transit crosses active regions which are much brighter than the rest of the stellar disk. If an active region is typically $\sim$~the size of a terrestrial planet, it might be completely occulted; as a result, whereas optical transits have depths from parts-per-million to the percent level, transits of active regions might have radiometric transit depths of percent or tens of percent, a very large signal that is likely to be easy to detect provided the star itself is detectable. 

One of the difficulties of detecting a transiting planet around a star at radio wavelengths is disentangling variable stellar emission from the impact of the planet. 
Radio emission from the Sun, particularly from active regions, can come in bursts of various types \citep{1947Natur.160..256P,1949AuSRA...2..214P} associated with coherent electron cyclotron emission \citep{1980SSRv...26....3M,1982ApJ...259..844M}. $\epsilon$~Eri, our prototype extrasolar radio star, was detected from 4-18~GHz in quiescent emission by \citet{2018ApJ...857..133B}, but in the 2-4~GHz band showed similar flux in a single burst. 

We therefore generically expect the radio emission of distant stars, especially from starspots, to vary rapidly with time. This may pose a problem for detecting scintillation or lensing as in Section~\ref{mag} in short cadence time series observations. We nevertheless expect the geometric transit to be robust, simply because it causes a strong dip with a characteristic shape in an otherwise variable time series, whose periodic ephemeris can be separately determined by optical studies. We therefore believe that while intrinsic stellar variability is likely to preclude radiometric transit surveys from being easily used to detect new planets, given known planets and models of the out of transit variability, the effects discussed in this paper are observable.

\subsection{Disentangling Stellar Activity from Transit Spectroscopy}
\label{transitspec}

Transmission spectroscopy of exoplanet atmospheres is achievable by measuring the differential transit depth as a function of wavelength. Where the atmosphere is more opaque, or more strongly scattering, the planet will appear larger and have a correspondingly deeper transit. Heterogeneity in the stellar photosphere, for instance from spots and faculae, can cause systematic errors in the inferred transit spectrum \citep{2017ApJ...834..151R,2018arXiv180308708A}, as it is necessary to know the distribution both of spots along the transit chord and also unocculted spots. This `transit light source problem' \citep{2017LPICo2042.4032R} is likely to be one of the major limitations on atmospheric studies with future large telescopes. The presence even of unocculted spots can bias transit depth measurements in white light and as a function of wavelength, and while this can be to some extent corrected by modelling the transit shape and ingress and egress durations \citep{2018arXiv180704886M}, it would be advantageous to have an independent measure of unocculted active region coverage. The radio spectrum of $\alpha$~Cen~AB has been used to constrain the spot coverage fraction \citep{2018arXiv180308338T}; with future large observatories, this will be possible for many more stars. As SKA1 (first light $\sim 2023$) and the five year mission of JWST (launch scheduled for 2021 at the time of writing) are likely to overlap in time, we recommend that JWST transit spectroscopy measurements be accompanied by simultaneous observations with SKA1 where possible to help constrain the transit light source effect, even in cases where SKA1 will lack the sensitivity to detect the transit itself. At higher sensitivities with SKA2, the transit depth in radio emission is related to the fraction of the radio flux contributed by the occulted active regions, permitting a more model-independent constraint on the spot distribution. Simultaneous transit spectroscopy and radio monitoring for the nearest solar-like stars or M~dwarf hosts will break degeneracies in the transit spectrum model related to stellar activity to obtain uncontaminated atmospheric transmission spectra.

\subsection{Cool Dwarfs}
\label{mdwarfs}

Planets, including habitable planets, are thought to be common around the low-mass M~dwarf stars \citep{2013ApJ...767...95D}. These include the nearest exoplanet to us \citep[Proxima~b:][]{2016Natur.536..437A} and the remarkable resonant chain of planets around TRAPPIST-1 \citep{2017Natur.542..456G,2017NatAs...1E.129L}. Many of these stars are known to be bright and variable in the radio \citep{2007ApJ...668L.163L,2008ApJ...673.1080B,2013A&A...549A.131A,2017arXiv170704264W}, and are in fact over-luminous in the radio relative to X-rays compared to earlier-type stars \citep{2012ApJ...746...23M}, lying above the G{\"u}del-Benz relation connecting X-ray and radio luminosity \citep{1993ApJ...405L..63G,2010ARA&A..48..241B}. Their rotation rates span a much wider range than more massive field stars \citep{2016ApJ...821...93N}, and the more rapidly rotating M~dwarfs show more signs of activity in H$\alpha$ emission \citep{2017ApJ...834...85N}. The fraction of M~dwarfs which are magnetically active, using $\text{H}\alpha$ emission as a proxy for magnetic activity, rises steeply from a few percent at M3 to $\sim 90\%$ at L0 \citep{2015AJ....149..158S}, which is possibly related to the onset of full convection around $\sim$\,M4 spectral type. Considering `ultracool dwarfs' later than this, and including L~dwarfs (brown dwarfs), around $5-10\%$ of ultracool dwarfs are found to flare in the radio \citep{2016ApJ...830...85R}. These flares are typically highly circularly polarized, while unpolarized, variable quiescent emission is also frequently detected \citep{2016MNRAS.457.1224L}. This emission likely arises from the electron cyclotron maser instability \citep{2008ApJ...684..644H,2018ApJ...854....7L}. The optical detection of bright regions on the surface of TRAPPIST-1 \citep{2018ApJ...857...39M} indicates the importance of stellar activity for understanding M~dwarf planets. 

The radio over-luminosity of some late-type dwarfs, combined with their very small radii making geometric transit depths very deep ($\sim 10\%$ or more for hot Jupiters in some cases) makes them ideal for transit observations as proposed in this work. The increase in sensitivity from the SKA and especially SKA2 will very significantly expand the volume and completeness of the sample of radio-loud ultracool dwarfs: from population synthesis modeling, so far only about 5\% of the radio-flaring ultracool dwarfs within 25~pc are thought to have been detected so far \citep{2017ApJ...845...66R}. The details of the magnetic dynamos of late type stars, particularly fully convective stars, are poorly understood, and they can give rise to magnetic field topologies with large scale dipole structure or messy small scale structure \citep{2008MNRAS.390..545D,2008MNRAS.390..567M,2010MNRAS.407.2269M}, with radio luminosities spanning more two orders of magnitude even for stars of the same spectral type \citep{2013A&A...549A.131A}. 

This variability and diversity means that on the one hand, stars such as LHS~3003 (M7, $d = 6.56 \pm 0.15$~pc) can be very luminous in the radio \citep[$0.270 \pm .04$~mJy:][]{2005ApJ...626..486B}, though variable to the extent that it was subsequently not detected by \citet{2009ApJ...700.1750O}. On the other hand, a search from radio emission from TRAPPIST-1 (M8, 12~pc) has recently placed stringent upper limits of $8.1 \mu$Jy across the $4-8$~GHz band, such that detecting transits on TRAPPIST-1 with the measured upper limits would be difficult even with the SKA. Given the significant variability of M~dwarfs, however, it is possible that these observations were taken in a period of low activity and that in future TRAPPIST-1 may be detectable in the radio. 

We have indicated the position of LHS~3003 on the sensitivity diagram in Figure~\ref{sensitivities}, pushing it out to 10~pc and 25~pc and considering the sensitivity needed to detect a 1\% transit. The very high radio luminosity from \citet{2012ApJ...746...23M} indicates that a 1\% transit would be detectable out to 25~pc. \citet{2018arXiv180406982B} have calculated that there are tens of thousands of M~dwarfs observable by TESS. \citet{2018AJ....155..180M} show that of cool dwarfs (including late~K~stars), selecting the 10,000 `easiest' targets for TESS out of 1,080,005 stars considered is likely to yield $\sim 133$ transiting planets. A comparison of \gaia~DR2 and SDSS data indicate there are at least $\sim 600$ M~dwarfs (Kiman et al., in prep.) within 25~pc, which Figure~\ref{sensitivities} shows may have transits detectable in the radio. For very late stars which are optically faint, their radio brightness may actually permit new exoplanet discoveries with the SKA, as opposed to the case for brighter stars where the radio approach will be primarily useful for following up known planets. 

The main limitation may be the very significant variability of these sources, both from rotational modulation and flares, which is likely to pose a significant challenge to transit detection in the radio as it does in the optical \citep{2016ApJ...829L..31D,2017AJ....153...93K}.  \citet{2018arXiv181000855V} find in a survey of~5 active M~dwarfs 22 bright, circularly polarized bursts on timescales of seconds to hours, inferring a duty cycle of $\sim 25\%$ at 1-1.4~GHz frequencies. This indicates that it will be impossible to straightforwardly detect planetary transits in this context without additional prior knowledge. Nevertheless, for a planetary ephemeris known from optical studies, transits will manifest themselves as  anomalous distribution of flare amplitudes - from plasma effects or occultation - and this will have the same utility as a transit across a steadily emitting star. Considering the extreme case of a light curve entirely consisting of flares with no detectable quiescent flux, even if the radio emission consists entirely of flares, you can measure the distribution of their amplitudes in and out of transit. 

Transits even of rocky planets may nevertheless be vital in establishing via transit tomography the radio emission distribution on the stellar surfaces, as the only method likely to directly spatially resolve this in the near future. We consider modelling  of M~dwarf transits to be beyond the scope of this paper: with complicated small-scale magnetic structure, beamed, coherent and polarized emission, extreme variability and bursty behaviour, it would take a very detailed analysis to predict the properties of transit light curves in details and predict their ultimate value. A promising next avenue is the method of \citet{2018ApJ...854....7L} to predict the radio emission and its properties based on magnetic field maps of the stellar surface from spectropolarimetry and from X-ray observations. Generating such radio maps and simulating transits would be a promising avenue for future work. It will be important to establish whether simulated input spot maps can be retrieved from radio transit light light curves, and if so, whether this information is complementary to or superior to what can be done with spectropolarimetry.

While small planets are common around M~dwarfs \citep{2013ApJ...767...95D}, these are unlikely to yield magnetospheric lensing detections. Only a handful of hot~Jupiters are known: \emph{Kepler}-45~b \citep[\gaia DR2: $381.95 \pm 6.38$~pc]{2012AJ....143..111J}, HATS-6~b \citep[\gaia DR2: $169.32 \pm 1.06$~pc]{2015AJ....149..166H}, and NGTS-1~b \citep[\gaia DR2: $218.12 \pm 1.04$~pc]{2018MNRAS.475.4467B}, although for NGTS-1b the host star does not appear to be active. At their distances and the brightness of LHS~3003, transits would not be detectable even with SKA2-Mid.

\section{Radio Transits Illuminating Planetary Magnetospheres}
\label{mag}

Radio observations can not only complement optical transits, but also enable qualitatively new science, probing planetary plasma environments. During a planetary transit, stellar emissions of all wavelengths are obstructed by the planet. 
In addition to this, stellar emissions also propagate through the atmosphere, ionosphere, and magnetosphere of the planet. 
For optical transit observations, atmospheric effects can be important, but ionospheric and magnetospheric effects are negligible.
By contrast, for radio transit observations, the effects of ionospheric and magnetospheric plasma may be important in the case of sufficiently dense environments.
Here we consider these effects, which occur in addition to the purely geometric planetary transit effect addressed in Section~\ref{geometrictransit}. Since magnetospheric lengthscales are on the order of planetary radii or greater, whereas ionospheric lengthscales are much smaller, we focus on magnetospheric effects. We note that because of the increased effective planetary radius if the magnetosphere is included, planets with grazing transits or which do not transit may be detected in the radio but not in the optical band.

Radio waves that propagate through this ionized medium will be affected in several different ways which may imprint information about magnetospheric properties on radio observations of the transit. 
\citet[hereafter WV17]{withers2017} considered four different aspects of a radio
signal propagating through a planetary environment (in the context of observing background radio point sources through Jupiter's magnetosphere): polarization,
frequency, power, and timing. Changes in polarization, frequency, and
timing are not applicable here because starspot radio emission is not
linearly polarized (though it can be circularly polarized), does not have a narrow and stable frequency, and is
not pulsed. For power, WV17 considered refractive
lensing in the context of smooth changes in refractive index with
radial distance. They applied that concept to the neutral atmosphere,
where the underlying assumption is well-satisfied. 
Here we consider refractive lensing caused by smooth changes in refractive index with
radial distance and also scintillation caused by small-scale irregularities in refractive index. In both scenarios, magnetospheric plasma density is the crucial physical property.

In this section we consider two potential transiting planets:
a Jupiter analogue at 1~AU and a hot Jupiter analogous to HD~189733~b.
Jupiter offers an established Solar System example of a particularly significant and dense magnetosphere.
The magnetospheres of hot Jupiters are not as well-characterized, but initial observations suggest that their magnetospheres can be substantial.
We define representative plasma density distributions for these two objects.
For the 1AU Jupiter analogue, we adopt the idealized model of WV17 in which the electron density $N$ effectively varies exponentially with radial distance.
Moving inwards from the boundary of the magnetosphere, the plasma scale height is 
40 $R_{J}$ between 100 $R_{J}$ and 28.6 $R_{J}$ (outer magnetosphere), 
2.5 $R_{J}$ between 28.6 $R_{J}$ and 10 $R_{J}$ (inner magnetosphere), 
and 4 $R_{J}$ between 10 $R_{J}$ and 5300 km above 1 $R_{J}$ ($\approx$1.07 $R_{J}$) (ad hoc transition region).

We neglect 
the ionosphere in this project. Here multipath effects cause complicated changes in received intensity, which are beyond the scope of the current exploratory study, and electron density does not vary exponentially with altitude, which precludes useful simplifying assumptions.
For a hot Jupiter, we develop a spherically-symmetric model of HD~189733b using parameter values from \citet{2013ApJ...773...62P}.
Plasma density $N$ in the exoplanetary magnetosphere is $7 \times 10^{10}$ cm$^{-3}$ at a radial distance of 1.75 $R_{P}$, where the planetary radius $R_{P}$ is 1.138 $R_{J}$. The plasma density $N$ decreases exponentially with increasing radial distance with scale height $H$ of 5000 km.
Although Jupiter orbits at 5 AU from the Sun, we assume that our Jupiter analogue orbits at 1 AU from its parent star.
Our hot Jupiter model orbits at 0.03 AU from its parent star like HD~189733b. In the case of the hot Jupiter in particular we stress that this model is highly idealised, using a Jupiter-like profile with higher density to a much closer orbit where we would expect a bow shock and an extended magnetotail, in order to make the problem straightforwardly tractable in the weak lensing and scintillation regime. Therefore in the case of the hot Jupiter the results presented below should be taken as qualitative and heuristic, and a more realistic prediction would require a~3D ray-tracing model.

\subsection{Lensing}

The mean refractive index profile of the exoplanetary magnetosphere will
cause refraction of the radio signal that passes through it. Due to this lensing, focusing or defocusing will occur, 
which will cause observed changes in received power that depend on the radial distance of
closest approach of the ray path to the exoplanet and other factors.

We consider the normalized intensity, defined as the ratio of the observed intensity $I$ to the intensity $I_{0}$ that would have been observed in the absence of refraction during the propagation of the radio signal through the planetary environment, given by Equation~6 in WV17.  

Large changes in intensity are caused by large planet to star separation $D$, large distance of closest approach between the ray and the planet $r$, high electron number densities $N$, and low frequencies $f$. 

The effects of lensing by a Jupiter analogue at 1 AU on received power at 0.1 GHz as predicted by WV17 are negligible: the normalized intensity is within 1\% of 1 at all distances. Effects on normalized intensity are even smaller at higher frequencies.  

\begin{figure}
\centering
\noindent\includegraphics[width=8cm,keepaspectratio]{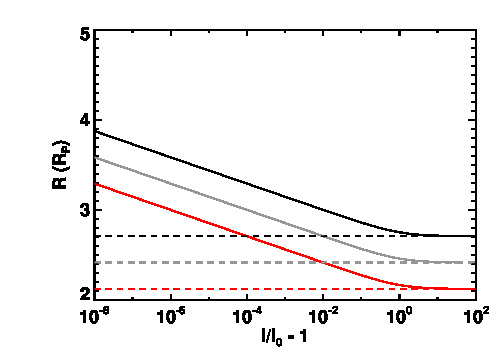}
\caption{
Solid lines show the dependence of normalized intensity, $I/I_{0}$, on distance of closest
approach of the ray path to the hot Jupiter. Black line is for 0.1 GHz, grey line is for 1 GHz, and red line is for 10 GHz. 
At small impact parameters at all frequencies weak lensing models predict increases in brightness by factors $\gg$~unity, indicating strong lensing effects and the breakdown of the weak lensing regime.
The colored dashed lines mark the radial distance at which the method used to calculate these results breaks down.
}
\label{fig:hotjupdefocusing}
\end{figure}

On the other hand, the effects of lensing by our hot Jupiter model at 0.03 AU on received power at several frequencies show much stronger effects (Figure~\ref{fig:hotjupdefocusing}). 
At large radial distances, $I$ is slightly greater than $I_{0}$, indicating an increase in intensity due to refractive lensing.
As radial distance decreases, $I/I_{0}$ increases until a critical transition is reached.

Changes in $\left(I/I_{0}\right)$ due to refractive lensing are predicted to be much greater for the hot Jupiter than for a Jupiter analogue exoplanet at 1 AU distance from its parent star. These differences can be explained in that the distance $D$ is approximately two orders of magnitude smaller for the hot Jupiter than the Jupiter analogue.
The scale height $H$ is approximately two orders of magnitude smaller;
the plasma density $N$ at 1.75 $R_{P}$ is approximately eight orders of magnitude larger;
and the differences in $H$ and $N$ cause greater changes in intensity for the hot Jupiter than the Jupiter analogue. 
The difference in $D$ has the opposite effect, but it is overwhelmed by the much more significant impact of the eight orders of magnitude difference in plasma density $N$.

These predictions are based on idealized assumptions about magnetospheric properties.
For instance, they neglect the high density bow shock in the magnetosphere of a hot~Jupiter or the asymmetric shape and extended tail typical of magnetospheres. These features will make magnetospheric effects more practically observable as a distortion of the radio light curve before the beginning and after the end of the geometric transit.

\subsection{Scintillation}

The deterministic, broadband effect of lensing is complemented by random scattering in the planetary magnetosphere. As a radio signal propagates through an inhomogeneous plasma, phase delays introduced by the plasma cause amplitude variations. The resultant random fluctuations in received intensity are referred to as scintillations. The size of the intensity fluctuations is related to properties of the plasma density fluctuations. 

An extensive body of literature exists concerning the effects on the
power of a radio signal of propagation through a region of plasma
density fluctuations in the ionosphere \citep{1963JATP...25..339B,1982IEEEP..70..324Y}, interplanetary space \citep{1964Natur.203.1214H}, interstellar space \citep{1972ApL....12..193H,1975ApJ...201..532L,1977ARA&A..15..479R}, and intergalactic space \citep{2016Sci...354.1249R}. The same scintillation effect occurs at optical wavelengths, where Earth's inhomogeneous turbulent atmosphere causes stars to twinkle. Solar System planets, however, do not twinkle to the naked eye, because the scintillation across their disk is incoherent and averages out to a more stable intensity; likewise, the presence or absence of scintillation can be used to constrain the angular size of radio sources. If the length scale of the inhomogeneities in the exoplanetary magnetosphere is similar to or greater than the size of a typical starspot, we expect scintillation to be important; if these inhomogeneities are smaller, however, we expect them not to be. As a typical starspot is of order the same size as a planet in our Solar System or in extrasolar systems where spot-crossing events have been observed in optical transits, it is plausible that some, but not all, extrasolar transiting systems may exhibit radio scintillation. 
With this in mind, we now consider the case of an inhomogeneous magnetosphere, and what can be determined about its structure from the short-term variability of the received radio flux. 

The ratio of the standard deviation of intensity to the mean intensity
is called the scintillation index, $m$. For $m \ll 1$, it satisfies \citep{1997sid..book.....A}:

\begin{eqnarray}
\label{xxx}
m \approx \lambda r_e \sqrt{a L} (N_\text{rms})
\end{eqnarray}

\noindent where $\lambda$ is the radio wavelength, $r_e$ is the classical electron
radius $e^{2} / \left (4 \pi \epsilon_{0} m_{e} c^{2} \right ) = 2.8 \times 10^{-15}$~m, $a$ is the typical size scale of density irregularities, $L$ is the
length-scale over which scattering occurs, and $N_\text{rms}$ is the
root-mean-square variation in electron density.
Equation~\ref{xxx} is valid for weak scintillation, where $m \ll 1$ \citep{1997sid..book.....A}.
In the limit of strong scintillation, $m = 1$ \citep{tatarskii1971, 1975ApJ...201..532L}.
In the interests of simplicity, we use the weak scintillation expression of Equation~\ref{xxx} for all calculations, but any values of the scintillation index $m$ that exceed unity should be replaced by the strong scintillation limit $m=1$. 

We consider the scintillation caused by the magnetosphere of a distant exoplanet of a background point source, using the above hot Jupiter and Jupiter analogues. 

For the 1~AU Jupiter analogue, we consider two possible values of $a$:
10 $R_{J}$, which
is the width of the equatorial plasma sheet \citep{1981JGR....86.8385B,2005JGRA..110.7227K}, and 0.1 $R_{J}$, which is a representative lengthscale for small structures in a
major feature of Jupiter's magnetosphere, the Io plasma torus \citep{doi:10.1002/2016JA023447}. For
the scattering lengthscale $L$, we assume $L = \sqrt{R H}$, where $R$ is the radial
distance and $H$ is the scale height that describes the exponential
dependence of plasma density on radial distance. This expression for $L$
is the effective path length of the ray path through the densest region
of the magnetosphere
\citep{1999JGR...10426997H}. $H$ is taken from
the model of \citet{withers2017}; it is $2.5 R_{J}$ for the inner
magnetosphere within $28.6 R_{J}$. For $N_\text{rms}$, we assume that
$N_\text{rms} / N = 0.1$. 

In both the hot Jupiter and Jupiter model we have considered, at small impact parameters we have very significant scintillation, in fact leaving the weak scintillation regime for which these calculations are valid.

The effects at higher frequencies are smaller (Equation~\ref{xxx}), but in both cases appreciable scintillation is predicted for lines-of-sight that do not
graze the planet's atmosphere and ionosphere. Closest approach distances
on the order of 10 $R_{J}$ are sufficient to achieve strong scintillation.

For the large scale, $m$ exceeds 1 for impact parameters from $b < 19.7 R_{J}$ (100~MHz) to $b < 6.6 R_{J}$ (10~GHz). 
For the small scale, $m$ exceeds 1 for $b < 13.8 R_{J}$ (100~MHz) to $b < 6.1 R_{J}$ (1~GHz); while at 10 GHz, $m$ does not exceed 1 for magnetospheric lines-of-sight.
Recall that we use the weak scintillation expression of Equation~\ref{xxx} for all calculations, so predicted values of the scintillation index $m$ that exceed unity should be replaced by the strong scintillation limit $m=1$. 

For our hot Jupiter model, we assume that the typical size scale for density irregularities $a$ is $0.1 R_{P}$ and that the root-mean-square fluctuation in density is 10\%.
These adopted values are not directly constrained by observations. Instead, they are plausible exploratory assumptions.

The predicted scintillation index $m$ exceeds 1 and we leave the weak scintillation regime for closest approach distances of approximately 3 $R_{P}$ for all three frequencies considered here (0.1, 1, and 10 GHz).

The scintillation index is smaller for the hot Jupiter than for the Jupiter analogue at large radial distances, but $m$ is larger for the hot Jupiter than for the Jupiter analogue at small radial distances. This can be explained by consideration of Equation~\ref{xxx}.
The most significant term in this equation is the plasma density $N$. At small radial distances, $N$ is much greater for the hot Jupiter than for the Jupiter analogue, as discussed above. However, $H$ is approximately two orders of magnitude smaller for the hot Jupiter than the Jupiter analogue.
Hence the plasma density $N$ decreases much more rapidly with radial distance for the hot Jupiter than the Jupiter analogue.
Consequently, $N$ is much smaller for the hot Jupiter than for the Jupiter analogue at large radial distances.

Given that there is a significant range of impact parameters for which the weak-limit expression predicts $m \gtrsim 1$, we suggest that it will be important to model the strong-scintillation regime to explore what radio transit light curves can encode at small impact parameters.
In practice, we expect that several length scales are present, for example the magnetospheric current sheet and Io plasma torus which justified the $10 R_J$ and $0.1 R_J$ irregularity scales respectively, and that it may be possible to retrieve the parameters of a model incorporating these effects deterministically from a sufficiently high-resolution radio light curve. It has previously been suggested that radio emission modulation analogous to the Jupiter-Io interaction may be used to detect exomoons \citep{2014ApJ...791...25N}, and we suggest that such a detection may also be possible through scintillation measurements.

We have calculated effects of defocusing and scintillation separately for convenience, but in reality both affect the same quantity --- the observed intensity. In principle, it might be possible to separate these effects by interpreting the time-averaged intensity in terms of defocusing and fluctuations in intensity with time in terms of scintillation. Given that for our toy hot Jupiter model both refraction and scintillation leave the weak effect regime for impact parameters smaller than a few planetary radii, a crucial further step will be to conduct detailed beam propagation modelling through a 3D magnetosphere to predict the effects of lensing and scintillation in the strong regime in each case.

\section{Conclusions}

Radio observations of planets transiting starspots are a promising direction for exoplanetary and stellar science with future large radio telescopes. It will become possible to detect the radio emission of many main sequence stars with SKA1-Mid, and the final SKA2-Mid will permit transit observations around nearby active stars. We recommend investment in more detailed theoretical models of stellar radio emission and planetary magnetospheres, and full-polarization ray-tracing models of hot Jupiter magnetospheres illuminated by realistic stellar activity maps, whose predictions will be tested by radiometric transit observations in coming years. Radio transit science with the SKA will complement the James Webb Space Telescope's five year mission, currently planned to commence in 2021. Even without detecting transits, the SKA1 observations of emission from nearby solar-like stars will help measure signatures of stellar activity which would otherwise contaminate optical transmission spectra. Furthermore with SKA1 and even more so with SKA2 detailed studies of many nearby M~dwarfs and their planetary systems will be possible, which will help constrain both planetary magnetospheres and stellar activity, and their interaction, with its consequences for habitability. We encourage radio astronomers with access to existing instruments, in particular the VLA, MeerKAT and ASKAP, to investigate the Sun and nearby stars with a view to constraining these presently-undeveloped models, and considering what else the SKA can do for stellar and exoplanetary astrophysics. 

\section*{Acknowledgements}

We would like to thank Laurent Pueyo for his encouragement, and Suzanne Aigrain, Vinesh Maguire-Rajpaul, Jane Kaczmarek, Rocio Kiman, Daniella Bardalez Gagliuffi, David Hogg, Jim Davenport, and Brett Morris for their valuable comments.

This work was performed in part under contract with the Jet Propulsion Laboratory (JPL) funded by NASA through the Sagan Fellowship Program executed by the NASA Exoplanet Science Institute.

BP acknowledges being on the traditional territory of the Lenape Nations and recognizes that Manhattan continues to be the home to many Algonkian peoples. We give blessings and thanks to the Lenape people and Lenape Nations in recognition that we are carrying out this work on their indigenous homelands.

This project was developed in part at the 2018 NYC Gaia Sprint, hosted by the Center for Computational Astrophysics of the Flatiron Institute in New York City. We have made use of data from the European Space Agency (ESA) mission \gaia (\url{https://www.cosmos.esa.int/gaia}), processed by the \gaia Data Processing and Analysis Consortium (DPAC, \url{https://www.cosmos.esa.int/web/gaia/dpac/consortium}). Funding for the DPAC has been provided by national institutions, in particular the institutions participating in the \gaia Multilateral Agreement.

This research has made use of the SIMBAD database, operated at CDS, Strasbourg, France; NASA's Astrophysics Data System; and the ALMA Sensitivity Calculator in the ALMA Observing Tool by Mark Nicol. The Observing Tool is the software tool that will be used for the preparation of ALMA Observing Projects. The Tool is built by the Proposal and Observing Preparation System team within the ALMA Computing IPT.

This research has made use of the following software: \textsc{IPython} \citep{PER-GRA:2007}, SciPy \citep{scipy}, \textsc{matplotlib} \citep{Hunter:2007}, \textsc{Astropy} \citep{2013A&A...558A..33A}, and \textsc{NumPy} \citep{van2011numpy}.

\bibliographystyle{mnras}
\bibliography{ms}

\bsp	
\label{lastpage}

\end{document}